\documentclass[11pt,,fleqn,reqno]{article}
\usepackage{color}
\usepackage{amsmath}
\usepackage{amssymb}
\usepackage{sidecap}
\usepackage{subfigure}
\usepackage{graphicx}
\usepackage{bm}
\newcommand{\br}{\mathbf{r}}

\newcommand{\bsigma}{\mathbf{\sigma}}

\newcommand{\bs}{\mathbf{s}}

\newcommand{\gind}{g_{\rm ind}}
\newcommand{\gpair}{g_{\rm pair}}

\newcommand{\calJ}{{\cal J}}
\newcommand{\calK}{{\cal K}}
\newcommand{\calH}{{\cal H}}

\newcommand{\blue}{\textcolor{blue}}

\newcommand{\order}{{\cal O}}
\newcommand{\tiC}{\tilde{C}}
\newcommand{\tim}{\tilde{m}}

\begin{document}

\bigskip

\bigskip

\bigskip

\ \\

\ \\

\bigskip

\bigskip

\begin{center}
{\LARGE \bf Statistical physics of pairwise probability models}

\bigskip

{\large Yasser Roudi $^1$, Erik Aurell $^{2,3}$, John A. Hertz $^{1,4}$}

\bigskip

$^1$NORDITA, Roslagstullsbacken 23, 10691 Stockholm, Sweden\\
$^2$Department of Computational Biology, AlbaNova University Centre, 106 91 Stockholm, Sweden\\
$^3$ACCESS Linnaeus Centre, KTH-Royal Institute of Technology, 100 44 Stockholm, Sweden\\
$^4$The Niels Bohr Institute, Copenhagen University, 2100 Copenhagen \O, Denmark\\
\end{center}

\bigskip

\bigskip

Statistical models for describing the probability distribution over the states of biological 
systems are commonly used for dimensional reduction. Among these 
models, pairwise models are very attractive in part because they can be fit using a reasonable amount of  data: knowledge of 
the means and correlations between pairs of elements in the system is sufficient. Not surprisingly, then, using pairwise models for studying
neural data has been the focus of many studies in recent years. In this paper, we describe how tools from statistical physics
can be employed for studying and using pairwise models. We build on our previous work on the subject
and study the relation between different methods for fitting these models and evaluating their quality.
In particular, using data from simulated cortical networks 
we study how the quality of various approximate methods for inferring the parameters in a pairwise model
depends on the time bin chosen for binning the data. We also study the effect of the size of the time bin on the model quality
itself, again using simulated data. We show that using finer time bins increases the
quality of the pairwise model. We offer new ways of deriving the expressions reported in our
previous work for assessing the quality of pairwise models.
\newpage

\section{Introduction}
In biological networks the collective dynamics of thousands to millions of interacting elements, 
generating complicated spatiotemporal structures, is fundamental for
the function. Until recently, our understanding of these structures was severely limited
by technical difficulties for simultaneous measurements from a large number of
elements. Recent technical developments, however, are making it more and more
common that experimentalists record data from larger and larger parts of the system.
A living example of this story is what is now happening in Neuroscience. Until
a few years ago, neurophysiology meant recording from a handful of neurons at a
time when studying early stages of sensory processing (e.g. the retina), and only
single neurons when studying advanced cortical areas (e.g. inferotemporal cortex). 
This limit is now rapidly going away, and people are collecting data from
larger and larger populations (see e.g. the contributions in \cite{Nicolesis07}). However, such data will not help us understand the system
{\em per se}. It will only give us large collections of numbers, and most probably we will
have the same problem that we had with the system, but now with the data set: we can
only look at small parts of it at a time, without really making use of its collective
structures. To use such data and make sense of it we need to find systematic ways
to build mathematically tractable descriptions of the data, descriptions with lower
dimensionality than the data itself, but involving relevant dimensions.

In recent years, binary pairwise models have attracted a lot of attention
as parametric models for studying the statistics of spike trains of neuronal populations \cite{Schneidman05,Shlens06,Tang08,Roudi09,Roudi09-2,Shlens09},
the statistics of natural images \cite{Bethge07}, and inferring neuronal functional connectivities \cite{Shlens06,Yu08}.
These models are the simplest parametric models that can be used to describe 
correlated neuronal firing. To build them one only needs the knowledge of the distribution of spike probability over pairs of 
neurons, and, therefore, these models can be fit using reasonable amounts of data. 
As is the case for any parametric model, one would like to know how useful a  pairwise
model is, that is, how well it can describe the statistics of spike trains and whether specific questions about the
network can be studied using the fitted model and its parameters. Furthermore, one would
like to have fast and reliable ways to fit the model. These issues can be naturally 
studied in the framework provided by statistical physics. Starting from a probability distribution over the states of a
number of elements, statistical physics deals with computing quantities such as correlation functions, entropy,
free energy, energy minima, etc., through exact or carefully developed approximate methods. It allows one to
give quantitative answers to questions such as: what is the entropy difference between real data and the parametric model? Is there a closed-form
equation relating model parameters to the  correlation functions? How well does a pairwise model approximate higher order statistics of the data? 

In this paper, after briefly reviewing the experimental results on pairwise models, we discuss the statistical physics approaches
that are useful for studying these models and what we find by employing them. We do this with the aim of providing a coherent 
framework for fitting and using pairwise models as well as evaluating their quality. We describe a number of approximate
methods for fitting pairwise models, their underlying assumptions, their relation to each other, and the physical intuition behind them.
We also study the quality of pairwise models to model the probability distribution over the spike trains. 
We have studied these issues in our previous work \cite{Roudi09,Roudi09-2}, and, here, we expand on this
work in three ways. We study how the quality of approximate inference methods described in \cite{Roudi09-2} 
and the quality of the pairwise models depend on the bin size chosen for binning the data. We describe
new ways of deriving the perturbative expansion we reported in \cite{Roudi09} as well as of
one approximation that we find to perform well (the TAP approximation). We also study how the 
probability of synchronous spikes according to the independent and pairwise models compare with the true probabilities. 

In the first part of the paper, we describe the naive mean-field approximation and the independent-pair approximation for fitting
the model parameters. When applied to binned spike trains, these approximations provide reliable estimates of the model parameters when
the size of the system and/or the time bins are small. We then describe how these approximations can be corrected to 
provide very accurate approximations for estimating the parameter even for large populations and large time bins:
the Thouless-Anderson-Palmer approximation and the Sessak-Monasson approximation. 

We then go on to study the quality of the pairwise model. As  briefly reviewed in sec. \ref{exp-results}, several experimental reports
showed that the pairwise models can provide perfect approximations to the true probability distribution of small populations 
of cells. Following this experimental work, Roudi et al \cite{Roudi09} performed a theoretical analysis of the pairwise models
to understand if their success on small populations can be extended to real-sized system.
To do this, they derived equations for the entropy difference between the pairwise model and the data,
as well as between an independent-neuron model and the data. This was done in the regime where the average number of spikes
generated in a time bin by the whole population is small compared to one, i.e. when $N\delta \ll 1$ where 
\begin{equation}
\delta=\frac{1}{N}\sum_{i} \nu_i \delta t,
\end{equation}
where $N$ is the number of neurons in the populations, $\nu_i$ is the firing rate of neuron $i$ and $\delta t$
is the size of the time bins. We denote this regime ''the low-rate regime'' (also called "the perturbative regime" \cite{Roudi09}), however, we emphasize that it is really the quantity
$N\delta$ that defines this regime, and not only the firing rate. In \cite{Roudi09}, the authors showed that in the low-rate regime,
most of the difference between the true entropy and the entropy of the independent-neuron model can be explained
by the pairwise model. However, this fact happens regardless of whether the true distribution for large $N$
can be well approximated by the pairwise model. In other words,
observing a good pairwise model in the low-rate regime does not tell us
if the pairwise model will be a good model for the real sized system. 

A crucial step in the derivation of entropy differences in  \cite{Roudi09} was to express the pairwise distribution in 
the so called Sarmanov-Lancaster representation \cite{Sarmonov63,Lancaster58}. Here we show that one can derive the same
expressions by approximating the partition function of a Gibbs distribution (Eq.~\eqref{pairwise_dist} in sec. \ref{sec:model-def}).
In addition to recovering the low-rate expansion of the entropies, we use the results to find the difference between
the probability of synchronous spikes according to the true model, the pairwise model and the independent model.
We also show that one can derive the results of \cite{Roudi09} by extending
the idea behind the independent-pair approximation to triplets of neuron, i.e. the 
independent-triplet approximation.   By taking the limit $\nu_i \delta t\to 0$ of the triplet approximation to the entropy of a given distribution, we 
show that the results of \cite{Roudi09} can be recovered in a considerably simpler way.

An important step in using binary pairwise models is making a binary representation 
of the spike trains. This is done by binning the spike trains into small time bins and assigning 
zero or one to each bin depending on whether there is a spike in it or not.  In fact, it was 
predicted in  \cite{Roudi09} that using finer and finer time bins, improves
the quality of the pairwise model. In this work we study the influence of the bin size on the quality of the 
approximate methods of fitting as well as the quality of the model.

Finally, we discuss two possible extensions of the binary pairwise models and mention a number of important
questions that they raise. We first describe the extensions to non-binary variables useful for studying
the statistics of modular models of cortical networks. We then describe an extension of the pairwise model to a model with asymmetric connections
which gives promising results for discovering the synaptic connectivity from neural spike trains. 

\subsection{The binary pairwise model}
\label{sec:model-def}
In a binary pairwise model, starting from the spikes recorded from $N$ neurons, one
first divides the spike trains into small time bins. One then builds a binary representation of the spike
trains  by assigining a binary spin variable $s_i(t)$ to each neuron $i$ and each time bin $t$, with
$s_i(t)=-1$ if neuron $i$ has not emitted any spikes in that time bin, and $s_i(t)=1$ if it
has emitted one spike or more. From this binary representation, the means and correlations between the neurons are
computed as
\begin{subequations}
\begin{align}
& \langle s_i \rangle_{\rm data}\equiv\frac{1}{T}\sum_{t} s_i(t)\\
&\langle s_i s_j \rangle_{\rm data}\equiv\frac{1}{T}\sum_{t} s_i(t) s_j(t)
\end{align}
\label{data_moments}
\end{subequations}
where $T$ is the total number of time bins.

The binary pairwise model of the data is then built by considering the following distribution over a set of $N$ binary 
variable $\bs=(s_1,s_2\dots,s_N)$
\begin{equation}
p_{\rm pair}(\bs)=\frac{1}{Z}\exp\left[\sum_i h_i s_i+\sum_{i<j} J_{ij} s_i s_j \right],
\label{pairwise_dist}
\end{equation}
and chosing the parameters, $h_i$ and $J_{ij}$ such that the means and pairwise correlations under this distribution matches those of the data 
defined in Eq.~\eqref{data_moments}, that is
\begin{subequations}
\begin{align}
&\langle s_i \rangle_{\rm pair}\equiv \sum_{\bs} p_{\rm pair}(\bs)\ s_i=\langle s_i \rangle_{\rm data}\\
&\langle s_i s_j \rangle_{\rm pair}\equiv \sum_{\bs} p_{\rm pair}(\bs)\ s_is_j=\langle s_i s_j \rangle_{\rm data}.
\end{align}
\label{moments_match}
\end{subequations}
Following statistical physics terminology, the parameters $h_i$ are usually called the {\em external fields} and $J_{ij}$, the {\em pairwise couplings}
or {\em pairwise interactions}. One important property of the pairwise distributions in Eq.~\eqref{pairwise_dist} is that it has the maximum amount of entropy
among all the distribution that have the same mean and pairwise correlations as the data, and it is thus usually called
the maximum entropy pairwise model. It is also called the Ising model, in accordance with the Ising model introduced in statistical
physics as a simple model of magnetic materials.

Although typically written in terms of $\pm 1$ spin variables, it is sometimes useful to write the pairwise distribution
of Eq.~\eqref{pairwise_dist} in terms of boolean variables $r_i=(s_i+1)/2$. 
In the rest of the paper, we sometimes use such boolean representation as some of the calculations and equations
become considerably simpler in this representation. We denote the external fields and the couplings
in the boolean representations by $\calH_{i}$ and $\calJ_{ij}$.

The external fields and pairwise couplings can be found using both exact and approximate methods as discussed in details in \cite{Roudi09-2} and reviewed
briefly in the following sections. Before reviewing these methods, we first review the experimental studies that have used the maximum entropy pairwise 
model to study the statistics of spike trains. 

\section{Review of experimental results}
\label{exp-results}
The maximum entropy pairwise model as a model for describing the statistics of neural
firing patterns was introduced in \cite{Schneidman05} and \cite{Shlens06}.
This work used the Ising model to study the response of retinal ganglion cells to 
natural movies \cite{Schneidman05}, steady spatially uniform illumination and 
white noise \cite{Shlens06}, as well as the spontaneous activity 
of cultured cortical networks \cite{Schneidman05}. The main goal of these studies was to find out how 
close the fitted pairwise model is to the true experimentally computed distribution over $\bs$.

As a measure of distance, these studies compared the entropy difference between 
the true distribution and the pairwise model and compared it with the entropy difference between the
the true distribution and an independent model. The independent
model is a distribution that has the same mean as the data but assumes the firing of
each neuron is independent from the rest, i.e.
\begin{subequations}
\begin{align}
&p_{\rm ind}(\bs)=\frac{1}{Z_{ind}}\prod_{i} \Big[\langle s_i \rangle_{\rm data} \delta_{s_i,1}+(1-\langle s_i \rangle_{\rm data}) \delta_{s_i,-1}\Big]=\frac{\exp\left[\sum h^{\rm ind}_i s_i\right]}{\prod_i 2\cosh(h^{\rm ind}_i)}.\\
&h^{\rm ind}_i=\tanh^{-1}(\langle s_i \rangle)
\end{align}
\end{subequations}
The measure of  misfit can thus be defined as
\begin{equation}
\Delta =\frac{\overline{S_{\rm pair}-S_{\rm true}}}{\overline{S_{\rm ind}-S_{\rm true}}},
\label{G-def}
\end{equation}
where the overline indicates averaging with respect
to many samples of $N$ neurons. The results of Schneidman et al showed that for populations of size $N=10$ or so, $\Delta$ was around 0.1. In order words, in terms of entropy difference,
the pairwise model offered a ten fold improvement over the independent model. In the other study, Shlens et al \cite{Shlens06} 
found $\Delta \sim 0.01$ for $N=7$. These authors also considered  a 
slightly different model in which only the pairwise correlation between the adjacent cells was used in the fit,
and correspondingly the pairwise interactions between non-adjacent cells were set to zero. The results
showed that this adjacent pairwise model also performed very well, with $\Delta \sim 0.02$ on average.
It is important to note that in Schneidman et al the stimulus induced long range correlations between cells, 
while in the data studied Shlens et al the correlations extended only to nearby cells. Following these studies, 
Tang et al \cite{Tang08} reproduced these observations to a large extent in other cortical preparations
and also concluded that the pairwise model can successfully approximate the multi-neuron firing patterns.
In a very recent study \cite{Shlens09}, Shlens et al extended their previous analysis to population of up to 
$100$ neurons concluding that the adjacent pairwise model performs very well even in this case,
with $\Delta \sim 0.01 - 0.02$. The studies of Shlens et al were done without stimulation or with white
noise stimulus, situations in which neurons do not exhibit strong long range correlations. It is 
still unclear how the pairwise model (not necessarily adjacent) will perform for cases in which
neuronal correlations exist between neurons separated by large distance\blue{,} i.e., when stimulated
by natural scenes.

The assumption that $\Delta$ is a good measure of distance rests upon
the assumption that we are interested in finding out how different
the true and model distributions are over the whole space of possible spike
patterns. This can be appreciated when we note that the definition in  Eq.~\eqref{G-def} is equivalent to
$\Delta = D_{KL}(p_{\rm true}||p_{\rm pair})/D_{KL}(p_{\rm true}||p_{\rm ind})$, where 
$D_{KL}(p||q)$ is the Kullback-Leibler divergence between two distribution $p(\bs)$ and $q(\bs)$
defined as $D_{KL}(p||q)=\sum_{\bs} p(\bs)\log(p(\bs)/q(\bs))$. Using this observation, we can
think of $\Delta$ as a weighted sum of the difference between 
the log probability of states according to the true distribution and according to the model distribution
normalized by the distance to the independent model.
Of course, we may not be interested in finding how different the two distribution are over the space of 
all possible states, but only how differently the model and true
distributions assign probabilities to a subset of {\em important} states. For instance,
in a particular setting, it may be important only to build a good model for the probability 
of all states in which some large number of neurons $M$ fire simultaneously (i.e., 
when $\sum_i r_i=M$), regardless of how different the two distribution are
on the rest of the states. It was found in \cite{Tkacik06} that
the pairwise model offers a significant improvement over the
independent model in modelling the experimental probability of synchronous multi-neuron
firing of up to $N=15$ neurons.

\section{Approximations for fitting binary pairwise models}

The commonly used Boltzmann Learning algorithm for fitting the parameters of the Ising model
is a very slow process, particularly for large $N$. Although effort has been made to speed up the 
Boltzmann learning algorithm \cite{Broderick07}, such modified Boltzmann algorithms
still require many gradient descent steps and long Monte Carlo runs for each step.
This fact motivates the development of fast techniques for calculating the model parameters
which do not rely on gradient descent or Monte Carlo sampling. A number of such approximations have been studied in \cite{Roudi09-2}. 
In what follows we describe these approximations, and in particular the relation between the 
simpler approximations (Naive mean-field approximation and the independent-pair approximation)
to the more advanced ones (TAP approximation and Sessak-Monasson approximation).

\subsection{Naive mean-field approximation (nMF) and the independent-pair approximations (IP)}

The simplest method \cite{Kappen98,Tanaka98} for finding the parameters of the Ising models from the data 
uses mean field theory:
\begin{equation}
\tanh^{-1}m_i = h_i + \sum_j J_{ij}m_j,		\label{nMFT}
\end{equation}
where $m_i=\langle s_i \rangle_{\rm data}$.  These equations express the effective field that determines 
the magnetization $m_i$ as the external field plus the sum of the influences of other
spins through their average values $m_j$, weighted by the couplings $J_{ij}$.  Differentiating with respect to a 
magnetization $m_j$ gives the inverse susceptibility (i.e., inverse correlation) matrix
\begin{equation}
(\mathbf{C}^{-1})_{ij} = - J_{ij}			\label{nMFT_corr}
\end{equation}
for $ i \neq j$, where  $C_{ij}=\langle s_i s_j \rangle_{\rm data}-m_i m_j$.  Thus, if one
knows the means $m_i$ and correlations $C_{ij}$ of the data, one can use Eq.~\eqref{nMFT_corr}
to find the $J_{ij}$ and then solve Eq.~\eqref{nMFT} to find the $h_i$. We call this approximation ``naive'' mean-field 
theory (abbreviated nMFT) to distinguish it from the TAP approximation described below, which
is also a mean-field theory.

In the independent-pair (IP) approximation, one solves the two-spin problem for neurons 
$i$ and $j$, ignoring the rest of the network. This yields the following 
expressions for the parameters in terms of the means and correlations:
\begin{subequations}
\begin{align}
&J^{\rm IP}_{ij}=\frac{1}{4}\log\left[\frac{((1+m_i)(1+m_j)+C_{ij}))((1-m_i)(1-m_j)+C_{ij})}{((1-m_i)(1+m_j)+C_{ij}))((1+m_i)(1-m_j)+C_{ij})}\right]\\
&h^{j}_i=\frac{1}{2}\log\left[\frac{(1+m_i)(1-m_j)-C_{ij}}{(1-m_i)(1-m_j)+C_{ij}}\right]+J_{ij}\label{hji}
\end{align}
\end{subequations}
where $h^{j}_i$ is the external field acting on $i$ when it is considered in a pair with $j$. It has been 
noted in \cite{Roudi09-2} that in the limit $m_i \rightarrow -1$ and $m_j\rightarrow -1$, $J^{\rm IP}_{ij}$ matches
the leading order of the low-rate expansion derived in \cite{Roudi09}.

Although the couplings found in the independent-pair approximation can be
directly used as an approximation to the true values of  $J_{ij}$,
relating the fields $h^{j}_i$ found from the independent-pair
approximation to those of the model is slightly tricky. The reason is that the expression 
we find depends on which $j$ we took to pair with neuron $i$. It is natural to think that we can
sum $h_i^j$ over $j$, i. e., over all possible pairings of cell $i$, to find the Ising model
parameter $h_i$. In doing so, however, we should be careful. The expression 
in Eq.~\eqref{hji} has two types of terms, those that only depend on $i$, i.e. the first terms in the
following decomposition, and those that involve $j$, i.e. the second and third terms below
\begin{equation}
h^{j}_i=\frac{1}{2}\log\left[\frac{1+m_i}{1-m_i}\right]+\frac{1}{2}\log\left[\frac{1-m_j-C_{ij}/(1+m_i)}{1-m_j+C_{ij}/(1-m_i)}\right]+J_{ij}.
\end{equation}
The first terms is the field that would have been acting on $i$ if it were not connected to any other neuron, and the rest are contributions 
from interactions with $j$. By simply summing $h^{j}_i$ over $j$, we will be overcounting this term, once for each pairing. In other words, the correct 
independent-pair approximation for $h_i$ will be
\begin{equation}
h^{\rm IP}_{i}=\frac{1}{2}\log \left[ \frac{1+m_i}{1-m_i} \right]+\frac{1}{2}\sum_{j\neq i} \log \left[ \frac{1-m_j-C_{ij}/(1+m_i)}{1-m_j+C_{ij}/(1-m_i)} \right ]+\sum_{j\neq i}J_{ij}.
\label{hpairs}
\end{equation}
Although simple in its derivation and intuition, in the limit $m_i\rightarrow -1$ for all $i$, Eq. \eqref{hpairs} recovers
both the leading term and the first order corrections of the low-rate expansion, as shown in Appendix \ref{lowrate-pairs}.

The simple naive mean-field and independent-pair approximations have been 
shown to perform well in deriving the parameters of the Ising model when the 
population size is small. 

In \cite{Roudi09-2}, we showed that for data binned at $10$ ms, the naive mean-field
and the IP approximations perform well in deriving the parameters of the Ising model when the 
population size is small. In Fig.\ \ref{fig:apps_prf} we extend this study
and evaluate how the quality of these approximations depend on the size of the time bin, $\delta t$. 
In this figure, we plot the $R^2$ value between the couplings found
from nMFT and IP approximations and the results of long Boltzmann runs as a 
function of the time bin chosen to bin the data. We do this for both $N=40$ and $N=100$. The simulations used to 
generate the spike trains and the Boltzmann learning procedure used in this figure are the same
as those reported in \cite{Roudi09-2}, with two exceptions. The first one is that here we use more gradient descent steps and
longer Monte Carlo runs, namely, 60000 gradient descent  steps and 40000 Monte Carlo steps per gradient descent step.
The second one is that here we use 10000 seconds worth of data for estimating the means and 
correlations. This is 2.5 times larger than what we used before. Both of these improvements were made to ensure
reliable estimates of the parameters, as well as the means and correlations, particularly for 
fine time bins. As can be seen in Fig.\ \ref{fig:apps_prf}, increasing either $N$ or $\delta t$ results in a decay in the quality of the IP approximation, as well as
its low-rate limit.  For the case of nMFT, a reasonable performance is observed only for $\delta t= 2$ms and $N=40$.
For large populations sizes and/or time bins nMFT is a bad approximation.
Given the strong dependence of the quality of these simple approximations 
on population size and the size of the time bin, we describe below how one can 
extend these approximations to obtain more accurate expressions for finding the external fields
and couplings of the pairwise model for large $N$ and $\delta t$.

\subsection{Extending the independent-pair approximation}
\label{IPA-EX}

\begin{SCfigure}
\includegraphics[height=11cm, width=7cm]{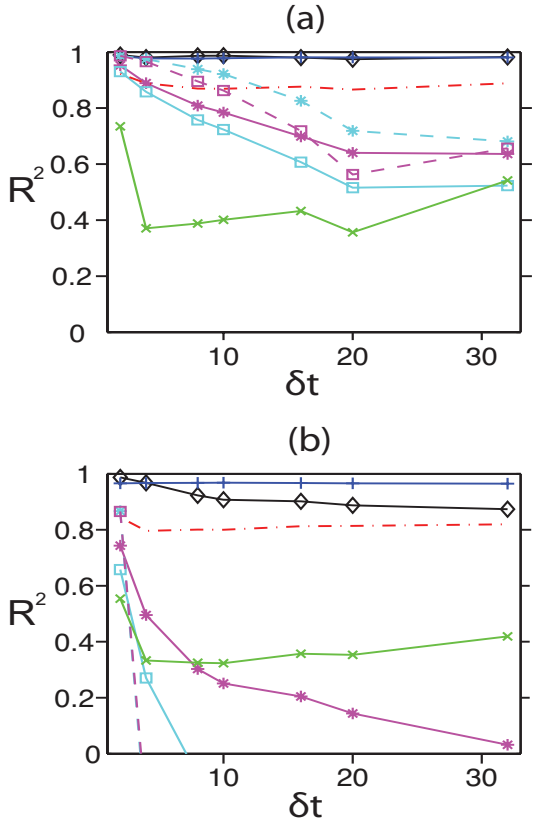}
\caption{The quality of various approximations for different time 
bins and populations sizes. Here we plot the $R^2$ values between
the couplings obtained using various approximate methods and
those found from Boltzmann learning versus $\delta t$ {\bf (a)} $N=40$ and {\bf (b)} $N=100$.
In both panel the colour code is as follows. Black, SM; Red, TAP; Blue, 
SM-TAP hybrid; Green, nMFT; Cyan, IP; Magenta low-rate limit of IP; 
Cyan with dashed line, IT; Megenta dashed lines, low rate limit of IT. 
For both population sizes and all time bins, the TAP, SM and hybrid approximations
perform very well.
\vspace{3cm}
}
\label{fig:apps_prf}
\end{SCfigure}

Extending the independent-pair approximation is in principle straightforward. Instead of 
solving the problem of two isolated spins, we can solve the problem of three spins, $i$, $j$, and $k$ as 
shown in Appendix \ref{triplet-expansion}. This will lead to the independent-triplet (IT) approximation. In Fig.\ \ref{fig:apps_prf}, we 
show the quality of the IT approximation for finding the couplings as compared to the Boltzmann solutions.
For $N=40$, we see that the IT approximation provides an improvement over IP for different values of $\delta t$.
In  Fig.\ \ref{fig:apps_prf}, we also looked at the quality of the low rate limit of the IT approximation. In Appendix \ref{triplet-expansion}
we show that, in the same way that the low rate limit of the IP approximation
gives us the leading order terms of the low-rate expansion in \cite{Roudi09},
the low rate limit of the IT approximation gives us the first order corrections to it. For $N=40$,
this is evident in the fact that the low-rate limit of IT outperforms the low-rate limit of
IP for $\delta t\le 15$ ms. When the population size is large, however, IT and its 
low rate limit outperform IP for only very fine time bins. Even for $\delta t=4$ ms, IP and IT and
their low rate limits perform very bad.

One can of course build on the idea of the IP and IT approximations and consider n=4, 5, \dots spins.
However, for any large value of $n$, this will be impractical and computationally expensive, for the following
reason. As described in Appendix \ref{triplet-expansion} for the case of the independent-triplet approximation, there
are two steps in building an independent-$n$-spin approximation. The first one is to express 
the probability of each of the $2^n$ possible states of a set of $n$ spins in terms of the means and 
correlations of these spins. This requires inverting the $2^n\times 2^n$ matrix. The second step, which is only present
for $n\ge 3$, is to express all correlations functions in terms of the means and pairwise correlations.
Both of these steps become exponentially hard as $n$ grows.  Nonetheless, as shown in Appendix \ref{triplet-expansion}, even 
going to the triplet level can be a very useful exercise, as it offers a new way of computing the difference between the 
entropy of the true model and that of the independent model (Eq.~\eqref{ents-perta} in sec. \ref{sec:ents}) as well as 
the difference between the entropy of the true model and that of the pairwise model (Eq.~\eqref{ents-pertb} \ref{sec:ents}). This 
derivation is considerably simpler than the original derivation of these equations based on the 
Sarmanov-Lancaster representation of the probability distribution described
in \cite{Roudi09} as well as the derivations in Appendix \ref{Ising-Low}, in which one starts by expanding
the partition function of a Gibbs distribution. Furthermore, the IT approximation also
yields a relation between the couplings and the means and pairwise correlations that
coincides with leading term and the corrections found by low-rate expansion, as shown in Appendix \ref{triplet-expansion}.

\label{Fig1}

\subsection{Extending the naive Mean-Field: TAP equations}
 
The naive mean-field, independent-pair and independent-triplet approximations are good for fitting the
model parameters when the typical number of spikes generated by  the whole population in a time bin 
is small compared to one, i.e., when $N\delta \ll 1$ \cite{Roudi09,Roudi09-2}.
For large and/or high firing rate populations, however, such approximations perform poorly for inferring the model
parameters. It is possible to make simple corrections to the naive mean-field approximation such that the resulting approximation performs well even
for large populations. This is the so called Thouless-Anderson-Palmer (TAP) approximation \cite{Thouless77}. The 
idea dates back to Onsager, who added corrections to the naive mean-field approximation, taking into account the effect of the
magnetization of a spin $i$ on itself via its influence on another spin $j$. Subsequently, 
it was shown that the resulting expression was exact for infinite-range spin-glass models \cite{Thouless77}.  The TAP equations are
\begin{equation}
\tanh^{-1}m_i= h_i+\sum_{j\neq i} J_{ij} m_j-\sum_{j\neq i} J^2_{ij} m_i (1-m^2_j).
\label{TAP}
\end{equation}
Differentiation with respect to $m_j$ then gives ($i \neq j$)
\begin{equation}
(\mathbf{C}^{-1})_{ij} = - J_{ij} - 2m_im_jJ_{ij}^2			\label{TAP_corr}
\end{equation}
One can solve Eqs.~(\ref{TAP_corr}) for the $J_{ij}$ and, after substituting the result in Eqns.~(\ref{TAP}), solve Eqns.~(\ref{TAP}) for the $h_i$ \cite{Tanaka98,Kappen98}.

There are several ways to derive this expression for the pairwise distribution Eq.~\eqref{pairwise_dist}. In Appendix \ref{TAP-BP}, we
show that these equations can also be derived from the celebrated 
Belief Propagation algorithm used in combinatorial optimization theory \cite{MezardMontanari2009}.
When applied to spike trains from populations of up to $200$ neurons, the inversion of 
TAP equations was shown to give remarkably accurate results \cite{Roudi09-2} for fitting the pairwise model.
In Fig.\ \ref{fig:TAPSM_scatter}, we show scatter plots comparing the couplings found by the TAP approximation versus the Boltzmann results for $N=100$
and $\delta t=2, 10, 32$ ms. The TAP approximations does well in all cases, and this is quantified in Fig.\ \ref{fig:apps_prf}.
In Fig.\ \ref{fig:apps_prf}, we demonstrate the power of inverting the TAP equations for inferring the couplings
for various time bins for both $N=40$ and $N=100$.

\subsection{Sessak-Monasson approximation (SM)}
\label{sec:SM}
Most recently, Sessak and Monasson \cite{Sessak09} developed a perturbative expansion
expressing the fields and couplings of the Ising distribution as a series expansion in the pairwise
correlations function $C_{ij}$. Some of the terms of the expression they found for the couplings 
could be summed up. It was noted in \cite{Roudi09-2}, that one can think of the resulting 
expression as a combination of the naive mean-field approximation and the independent-pair
approximation. The Sessak-Monasson result can be written as
\begin{equation}
J^{\rm SM}_{ij}=-(C^{-1})_{ij}+J^{\rm IP}_{ij}-\frac{C_{ij}}{(1-m^2_i)(1-m^2_j)-(C_{ij})^2},
\label{JSM}
\end{equation}
The reason why the last terms should be subtracted is discussed below. 

\begin{SCfigure}
\includegraphics[height=11 cm, width=7 cm]{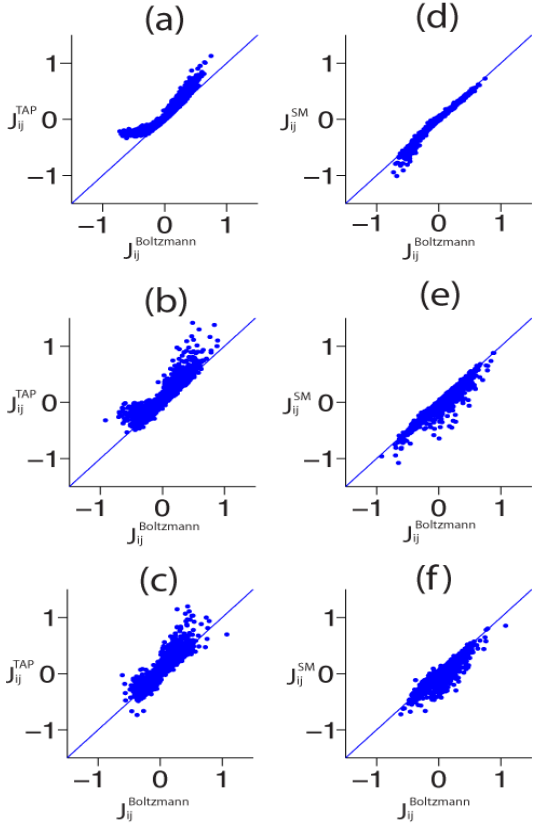}
\caption{Scatter plots showing the results of TAP and SM approximations versus the Boltzmann results for various
time bin sizes $\delta t$ and $N=100$. Panels {\bf (a)}, {\bf (b)}  and {\bf (c)} show the TAP results versus the Boltzmann results for
data binned at $2$ ms, $10$ ms and $32$ ms, respectively.  {\bf (d)}, {\bf (e)}  and {\bf (f)} show the same but for the SM approximation.
Note that the structure of the error in estimating the couplings from the TAP equations changes when the size 
of the time bin is increased.
 \vspace{4.5cm}
}
\label{fig:TAPSM_scatter}
\end{SCfigure}

Let us consider two neurons connected to each other. In the independent-pair 
approximation we calculate the fields and couplings for this pair exactly 
within the assumption that they do not affect the rest of the network and vice versa.  If we were 
to find the coupling between this pair of neurons using the naive mean-field 
approximation Eq.~\eqref{nMFT_corr}, the result would just be the last term in Eq.~\eqref{JSM}.
The reason why we should subtract it is now clear: the first term in Eq.~\eqref{JSM} includes
a naive mean-field solution to the pair problem. We subtract this part and replace it 
by the exact solution of the pair problem. Fig.\ \ref{fig:apps_prf}  shows that this result is very robust to changing $\delta t$, 
although for $N=100$, we can note a small decay in $R^2$ with $\delta t$. The good performance of the
SM approximation can be also seen in the scatter plots shown for $N=100$
in Fig.\ \ref{fig:TAPSM_scatter}. These observations support  
the SM approximation as a very powerful way of inferring the functional connections. Following the observation made in \cite{Roudi09},
in Fig.\ \ref{fig:apps_prf}, we also show how a simple averaging of the best approximate methods, i.e., TAP inversion and SM 
can provide a very accurate approximation to the couplings across different time bin and population sizes. 

\section{Assessment of model quality}

The experimental result that the binary pairwise models provide very good 
models for the statistics of spike trains is very intriguing. 
However, the message they carry about the architecture and function 
of the nervous system is not clear. This is largely due to the 
fact that, as reviewed in sec. \ref{exp-results}, the experimental studies were conducted on 
populations of small number of neurons ($N\sim 10$) and their implications on the
real sized system are not trivial. Is it the case that observing a very 
good pairwise model on a subsystem of a large system constrains the structure and function of the 
real sized network? Does it mean that there is something
unique about the role of pairwise interactions in the real sized system?
Answering this question depends to a large extent on answering {\em the extrapolation problem}: 
to what degree the experimentally reported success of pairwise models holds for the real sized system? 
In what follows, we discuss some theoretical results that bear on this question.

\begin{figure}[h]
\centering
\includegraphics[height=12 cm, width=14 cm]{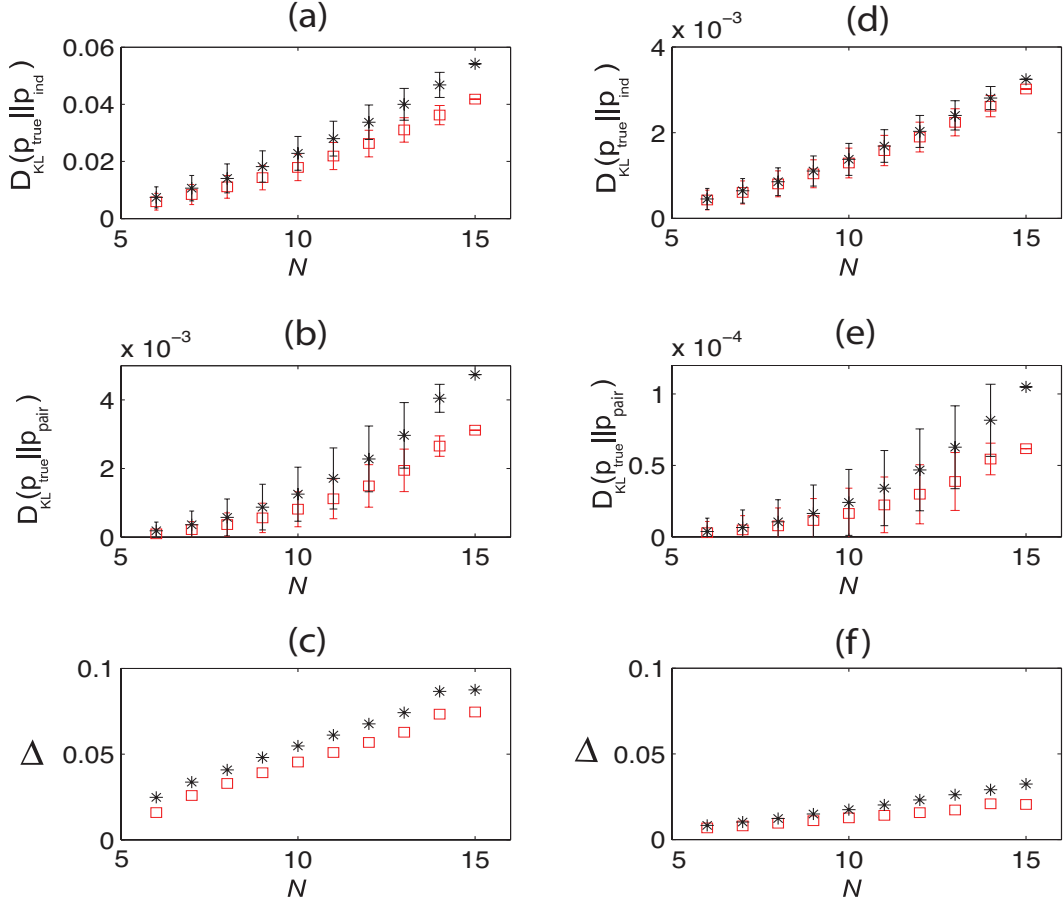}
\caption{The quality of the pairwise and independent models for different time bins and populations sizes. {\bf (a)}  $D_{KL}(p_{\rm true}|| p_{\rm ind}) $
versus $N$, {\bf (b)} $D_{KL}(p_{\rm true}|| p_{\rm pair})$ versus $N$ and  {\bf (c)} $\Delta$ versus $N$ all for $\delta t= 10$ ms.
{\bf (d)},  {\bf (e)},  {\bf (f)} show the same things for $\delta t= 2$ ms. In all panels, the black stars represent quantities as computed directly from 
the simulated data, while the red squares show the predictions of the low-rate expansion, i.e. Eqs.\ \eqref{ents-pert} and \eqref{G-pert}. 
We have used $18000$ seconds of simulated data for computing the plotted quantities and have corrected for 
finite sampling bias as described in \cite{Roudi09-2}.}
\label{fig:ents}
\end{figure}

\subsection{Entropy difference}
\label{sec:ents}
The extrapolation problem was first addressed in \cite{Roudi09} by analyzing the dependence of the 
misfit measure $\Delta$ (defined in Eq.~\eqref{G-def}) on $N$. The authors considered an 
arbitrary true distribution and computed the KL divergence between this distribution
and an independent-neuron model as well as between it 
and the pairwise model. This was done using a perturbative expension
in $N \delta \ll 1$. The results was the following equations:
\begin{subequations}
\begin{align}
&D_{KL}(p ||\ p_{\rm ind})=S_{\rm ind}-S_{\rm true}=\gind N (N-1) \delta^2+\order(N\delta^3) \label{ents-perta}\\
&D_{KL}(p ||\ p_{\rm pair})=S_{\rm pair}-S_{\rm true}=\gpair N (N-1)(N-2) \delta^3+\order((N\delta^4) \label{ents-pertb}
\end{align}
\label{ents-pert}
\end{subequations}
where $\gind$ and $\gpair$ are constant that do not depend on $N$ or $\delta t$
and are defined in Eqs.~\eqref{gind} and \eqref{gind}. Using 
these expression for the KL divergences yields
\begin{equation}
\Delta=\frac{\gpair}{\gind} (N-2) \delta.
\label{G-pert}
\end{equation}

Eqs.~\eqref{ents-pert} and \eqref{G-pert} show that for small $N \delta $, $\Delta$ will be 
very close to $0$,  independent of the structure of the true distribution. In other 
words, in this regime, a very good pairwise-model fit is a generic property and does not tell us 
anything new about the underlying structure of the true probability 
distribution. It is important to note that the perturbative expansion is always valid if 
$\delta t$ is small enough. That is, simply by choosing a sufficiently small time bin, we can 
push $\Delta$ as close to $0$ as we want.  

In \cite{Roudi09},  $\gpair$ and $\gind$ are related to the 
parameters of the pairwise model up to corrections of $\order(N\delta)$. In Appendix  \ref{Ising-Low}, we 
present a different derivation by expanding the partition function of a true Gibbs probability distribution 
around the partition function of a distribution without couplings.
In the following subsection, we also use the results of this derivation to compare the probability of synchronous spikes 
under the model and the true distributions.  Furthermore, in Appendix \ref{triplet-expansion}, we 
extend the idea beyond the independent-pair approximation and approximate the 
entropy of a given distribution as a sum over the entropies of triplets
of isolated neurons. We show that this approach leads to Eqs.~\eqref{ents-pert} in a
substantially simpler way than those reported in \cite{Roudi09}
and Appendix \ref{Ising-Low}.

In Fig.\ \ref{fig:ents}, we show how  $D_{KL}(p ||\ p_{\rm ind})$, $D_{KL}(p ||\ p_{\rm pair})$
and $\Delta$ vary with $N$ and $\delta t$ for data generated from
a simulated network. We have also plotted the predictions of the low-rate expansion 
Eqs.~\eqref{ents-pert} and \eqref{G-pert}. As shown in these figures, the low-rate expansion 
nicely predicts the behaviour of the measurements from the simulations particularly 
for small $N$ and $\delta t$.  For $\delta t= 10$ ms, we have
$\delta_{10} =0.076$ and for $\delta t=2$ ms, we have $\delta_{2}=0.019$
(note that this gives $\delta_{10} /\delta_{2}=3.95$, a ratio that would have been equal to $5$
if the bins were independent). Both the results from the low-rate expansion and those
found directly from the simulations show that using finer time bins decrease $\Delta$ for fixed $N$.

\subsection{Probability of simultaneous spikes}
\label{sec:synch-spikes}
As discussed in sec. \ref{exp-results}, in addition to entropic measures such 
as KL divergence and $\Delta$, which in a sense asks how well the model approximates the 
experimental probabilities of all possible spike
patterns, we can restrict our quality measure to a subset of possible spike patterns.  We 
can, for instance, ask how well the pairwise model 
approximates the probability of $M$ simultaneous spikes. Similar to the case 
of $\Delta$, before getting too impressed about the power of pairwise models 
in approximating the true distributions, we should find out what we expect in the 
case of an arbitrary, or random, true probability distribution.

Suppose now that we have a distribution over a set of variables of the form of Eq.~\eqref{def-third}. For this distribution,
the probability that a set of $M$ neurons, $I=\{i_1,i_2,\dots, i_M\}$ out of the whole population of $N$
fire in a time bin while the rest do not is
\begin{equation}
\log p_I= \sum_{i\in I} \calH_i+\sum_{i<j \in I} \calJ_{ij}+\sum_{i<j<k \in I} \calK_{ijk} +\dots -\log Z.
\end{equation}
Averaging over all possible $I$ we get
\begin{equation}
q(M,N)\equiv {N \choose M}^{-1}\sum_{I} \log p_I=M \overline{\calH}+{M \choose 2} \overline{\calJ} + {M \choose 3} \overline{\calK}+\dots-\log Z.
\end{equation}
where $\overline{\calH}$, $\overline{\calJ}$ and $\overline{\calK}$ are the means of the fields and pairwise, third-order etc 
coupling. In Appendix \ref{Ising-Low}, we show that, to leading order in $N \delta$, the external fields and 
pairwise couplings of fitted models (independent or pairwise) match those of the true
model (see Eqs.~\eqref{MH} and \eqref{CJ}). Using this, we see that

\begin{subequations}
\begin{align}
&q_{\rm true}(M,N)-q_{\rm ind}(M,N)\sim (\overline{\calJ} N) (M/N)^2+ (\overline{\calK} N^2) (M/N)^3 +\order((M/N)^4)\\
&q_{\rm true}(M,N)-q_{\rm pair}(M,N)\sim (\overline{\calK} N^2) (M/N)^3 +\order((M/N)^4).
\end{align}
\label{synch-spike}
\end{subequations}
To have well defined behviour in large $N$ limit,  one should have $\overline{\calJ} N\sim 1$ and  $\overline{\calK} N^2\sim 1$. Eq.~\eqref{synch-spike} show 
that, for $M/N \ll 1$, both the independent and pairwise models are close to the true distribution. For $M/N\sim \order(1)$ (of course still $M/N < 1$),
the difference between the model and true probabilities of observing 
$M$ synchronous spike increases. For all ranges of $M$, the difference is larger for the independent model. These predictions
are consistent with the experimental results found in the retina \cite{Tkacik06}.

In the above calculation, we compute the difference between the true and model values of $q$ 
the {\em mean of the log probability of $M$ synchronous spikes}. However, one can ask 
about the {\em log mean probability of $M$ synchronous spikes}, i.e.
\begin{equation}
w(M,N)\equiv -\log {N \choose M}+ \log \sum_{I}  \exp\left [\sum_{i\in I} \calH_i+\sum_{i<j \in I} \calJ_{ij}+\sum_{i<j<k \in I} \calK_{ijk} +\dots \right] - \log Z.
\end{equation}
Caculating $w$ is in theory very hard, because the second term above
involves calculating the averages of exponential functions of variables. However, if the population is 
homogenous enough, such that the average couplings and fields from one sample of $M$ neurons
to the next does not change much, we can approximate the average of the exponential of a variable 
with the exponential of the average it. Doing this, will again lead to Eq.~\eqref{synch-spike}. The difference
between the two measure $w$ and $q$ will most likely appear for small $M$ where the averages of the fields
and couplings depends on the sample of $M$ neurons more strongly than when $M$ is large.
 
\section{Extensions of the binary pairwise model}

In the previous sections, we described various approximate methods for fitting a pairwise model of the type
of Eq.~\eqref{pairwise_dist}. We also studied how good a model it will be for spike trains, using
analytical calculations and computer simulations. As we describe below, there are two issues with a model of
the type of Eq.~\eqref{pairwise_dist} that lead to new directions for extending the pairwise models studied here. 

The first issue is the use of binary variables as a representation of the states of the system.
For fine time bins and neural spike trains, the binary representation serves its purpose very well. 
However, in many other systems, a binary representation will be a naive simplification. Examples of
such systems are modular models the cortex in which the state of each cortical module
is describe by a variable taking a number of states usually much larger than 2. In the 
subsection \ref{nonbinary} we briefly describe a simple non-binary model useful for modelling the statistics of such systems.

The second issue is that by using Eq.~\eqref{pairwise_dist} in cortical networks, 
one is essentially approximating the statistics of a highly non-equilbium system with asymmetric
physical interactions, e.g. a balanced cortical network, by an equilibrium distribution with symmetric interactions. This manifests itself 
in a lack of a simple relationship between the functional connectives to real physical connections. In our 
simulations we observed that there was no obvious relation between the
synaptic connectivity and the inferred functional connections. Second, as we showed here
and in our previous work,  for large populations, the model quality decays. Although one can avoid this decay
by decreasing $\delta t$ as $N$ grows, eventually one will get into the regime of very fine $\delta t$,
where the assumption of independent bins used to build the model does not hold any more and one should
start including the state transitions in the spike patterns \cite{Roudi09}.
In fact, Tang et al \cite{Tang08} showed that even in the cases that the pairwise distribution of Eq.~\eqref{pairwise_dist}
is a good model for predicting the distribution of spike patterns, it will not be a good one for predicting the 
transition probabilities between them. These observations encourage one to go beyond an
equilibrium distribution with symmetric weights. In the second extension, described in subsection \ref{dynamics}, we propose one such model,
although a detailed study of the properties of such model is beyond the scope of this paper.

\subsection{Extension to non-binary variables}
\label{nonbinary}

The binary representation is probably a good one for spike trains binned into fine time bins. 
However, for larger time bins where there is a considerable probability of observing more than
one spike in a bin, as well as for a number of other systems, the binary representation may only serve 
as a naive simplification and going to non-binary representations is warranted. Example of 
such systems include the protein chains and modular cortical models. In probabilistic models
of protein chains, each site is represented by a non-binary variable that takes
one of its possible $q$ states depending on the amino acid that sits on that site. In a number of models
for the operations of cortical networks, one considers a network of interconnected modules, the state of each of
which is represented by a non-binary variable \cite{Kro+05,Russo08}. Each state of one such variable correspond to e.g. one of the many 
memory states stored in the corresponding module.

For a set of non-binary variables $\bsigma=(\sigma_1,\sigma_2,\dots,\sigma_N)$, $\sigma_i=1\dots q$, one can simply 
write down a maximum entropy pairwise Gibss distribution as
\begin{subequations}
\begin{align}
&p(\bsigma)=\frac{1}{Z}\exp\left[\sum_i \sum_{\alpha} \calH^{\alpha}_i u_{\alpha\sigma_i} + \sum_{i<j} \sum_{\alpha,\beta} \calJ^{\alpha \beta}_{ij} u_{\alpha\sigma_i} u_{\beta\sigma_j}\right],\\
&u_{\alpha\sigma_i}=\delta_{\alpha\sigma_i}-\delta_{1\sigma_i},\label{subt1}
\end{align}
\label{nonbin-def} 
\end{subequations} 
where $\alpha$ and $\beta$ go from $1\dots q$ and index the $q$ possible states of each
variable. For $q=2$, the above distribution reduces to the binary case with boolean variables,
and when one forces $\calJ^{\alpha \beta}_{ij}=0$ for $\alpha\neq \beta$ one recovers the $q-state$ Potts model.
Similar to the binary case, here also one is given the experimentally observed values 
of $\langle u_{\alpha\sigma_i}\rangle_{\rm data}$ and $\langle u_{\alpha\sigma_i} u_{\beta\sigma_j}\rangle_{\rm data}$
and wants to infer the fields and couplings that consistent with them. 

The approximate methods described in this paper can be adopted, with some effort, to the case of non-binary variable as well.
In particular, it is easy to derive the difference between the entropy of the true distribution, the pairwise model
and the independent model in the low-rate limit of the non-binary model. Assuming that $\langle u_{\alpha\sigma_i}\rangle_{\rm data}=\order(\epsilon)$ for $\alpha\neq 1$,
the low-rate regime in the case of non-binary variables is characterised by $N(q-1) \epsilon \ll 1$. In this regime, similar to the binary
case, $S_{\rm pair}-S_{\rm true} \propto (N(q-1) \epsilon)^3$ and $S_{\rm ind}-S_{\rm true} \propto (N(q-1) \epsilon)^2$ and $\Delta= N(q-1) \epsilon$,
and consequently the pairwise model performs very well in the low-rate regime.

As we described before, the low-rate regime is where most experimental 
studies on binary pairwise models were performed, and the result of the low-rate expansion of the entropies
explains the reported success of binary pairwise models in those studies. On the other hand, the low-rate regime of the
non-binary variable may be of little use. This is because the systems to which the non-binary representation
should be applied are unlikely to fall into the low-rate regime. For instance, in the case of modular memory networks
the low-rate regime would be the case in which the network spends a significantly larger time in its ground state (no memory retrieved)
compared to the time it spends operating and retrieving memory. A more likely scenario is the one where all the states (memory or no-memory) have
approximately similar probabilities of occurrence over a period of time. How useful pairwise models are
in describing the statistics of such non-binary systems away from the trivial regime of the low-rate expansion is not known.
Studying the quality of non-pairwise models in these cases and developing efficient ways to fit such models, in particular
based on extensions of the powerful approximations such TAP and SM to non-binary variables will be the focus of future work.
In particular, it is important to note that writing the SM approximation for the non-binary case will be a straightforward
task in light of the relation we described between the SM, nMFT and IP approximations in sec. \ref{sec:SM}

\subsection{Extension to dynamics and asymmetric interactions}
\label{dynamics}

The Glauber model \cite{Glauber63} is the simplest dynamical model that has a stationary
distribution equal to the Ising model distribution Eq.~(\ref{pairwise_dist}).  It is defined by a 
simple stochastic dynamics in which at each timestep $\delta t = \tau_0/N$ one spin is chosen 
randomly and updated, taking the value +1 (i.e., the neuron spikes) with probability 
\begin{equation}
p_+ = \frac{1}{1 + \exp [-2 (h_i + \sum_j J_{ij}s_j)]}.
\label{update_rule}
\end{equation}
Although the interactions $J_{ij}$ in the static Ising model are symmetric 
(any antisymmetric piece would cancel in computing the distribution (\ref{pairwise_dist})), a 
Glauber model with asymmetric $J_{ij}$ is perfectly possible \cite{Crisanti88,Ginzburg94}.  

This kinetic Ising model is closely related to another class of recently-studied network 
models called generalized linear models (GLMs)\cite{Truccolo05,Okatan05,Pillow08}.  In a GLM,  
neurons receive a net input from other neurons of the linear form, 
\begin{equation}
\sum_j \int_{0}^{\infty} d\tau J_{ij}(\tau)s_j(t-\tau)
\label{linear_kernel}
\end{equation}
and spike with a probability per unit time equal to a function $f$ of this input. 
Maximum-likelihood techniques have been developed for solving the
inverse problem for GLMs (finding the linear kernels that give the stimulus input
and that from the other neurons in the network, given spike train data) \cite{Truccolo05,Okatan05}. They
have been applied successfully to analyzing spatiotemporal correlations in 
populations of retinal ganglion cells \cite{Pillow08}.  

The Glauber model looks superficially like a GLM with instantaneous 
interactions and $f(x)$ equal to a logistic sigmoid function $1/(1+\exp(-2x))$.  But there is a
difference in the dynamics: spins are not spikes. In the Glauber model, a spin/neuron retains 
its value (+1 or $-$1) until it is chosen again for updating.  Since the updating is random, 
this persistence time is exponentially distributed with mean $\tau_0$.  Thus a 
Glauber-model ``spike'' has a variable width, and the autocorrelation function of 
a free spin exhibits  exponential decay with a time constant of $\tau_0$.  In the 
GLM, in contrast, a spike is really a spike and the autocorrelation function
(for constant input) is a delta-function at $t=0$.   

The time constant characterizing the kernel $J_{ij}(\tau)$ in a GLM has a similar 
effect to $\tau_0$ in the Glauber model, but a GLM with an exponential kernel 
is not exactly equivalent to the Glauber model.  In the GLM, the effect of a 
presynaptic spike is spread out and delayed in time by the kernel, but once it is 
felt, the postsynaptic firing rate changes immediately.  In the Glauber model, the 
presynaptic ``spike'' is felt instantaneously and without delay, but the firing state of 
the neuron takes on the order of $\tau_0$ to change in response.

GLMs grew out of a class of single-neuron models called LNP models.  The 
name LNP comes from the fact that there is a linear (L) filtering of the inputs, the 
result of which is fed to a nonlinear (N) function that specifies an instantaneous 
Poisson (P) firing rate.  In the earlier studies, the focus was on the sensory periphery, 
where the input was an externally specified ``stimulus''.  An aim of this modeling effort 
was to improve on classical linear receptive field models.  Thus, in the usual 
formulation of a GLM network, one writes the total input as a sum of two terms, one 
linear in the stimulus as in the LNP model and the other linear in the spike trains of
the other neurons.  Of course, one can trivially add a ``stimulus'' term in a Glauber model, so this
 differences is not an essential one.  

Similarly, interactions with temporal kernels $J_{ij}(\tau)$ can be included straightforwardly in a
Glauber model.  Such a model is equivalent to a GLM in the limit  $\tau_0 \to 0$. (One has to multiply
the kernels by $1/\tau$ while taking the limit, because the integrated strength of a "Glauber spike" is 
proportional to $\tau_0$, while that of an ordinary spike in a GLM is 1.)

One can derive a learning algorithm for a Glauber model, given its history, in a standard way, 
by maximizing the likelihood of the history.  The update rule, which is exact in the same 
way that Boltzmann learning is for the symmetric model, is
\begin{equation}
\delta J_{ij} = \eta \langle [ s_i(t_i+\epsilon)- \tanh (b_i + \sum_k J_{ik}\delta s_k(t_i)) ] \delta s_j(t_i) \rangle,
\label{Glauber_learn}
\end{equation}
where $\delta s_j(t) = s_j(t) -\langle s_j \rangle$, the average is over the
 times $t_i$ at which unit $i$ is updated, and $b_i = \tanh^{-1}\langle s_i \rangle = h_i + \sum_j J_{ij}\langle s_j\rangle$.

One can also get a simple and potentially useful approximate 
algorithm which requires no iteration by expanding the tanh in 
eqn.~(\ref{Glauber_learn}) to first order in the $J_{ij}$ (i.e., to first order around the
independent-neuron model).  Then at convergence ($\delta J_{ij} = 0$) we have
\begin{equation}
\langle \delta s_i(t_i+\epsilon)\delta s_j(t_i) \rangle = (1-\langle s_i \rangle^2)
\sum_k J_{ik} \langle \delta s_k(t_i) \delta s_j(t_i)\rangle,					\label{first_order_alg}
\end{equation}
which is a simple linear matrix equation that can be solved for the $J_{ij}$.
 

\section{Discussion}

The brain synthesizes higher-level concepts from multiple inputs received, and in many fields of research 
scientists are interested in inferring as simple a description as possible, given potentially vast amounts of
data. Such processes of learning take a set of average values, correlations,
or any other pattern in the data, and from these arrive at another representation, 
which is useful for speed or accuracy of prediction, for 
compressed storage of the data, as a grounds for decision making, or 
in any other aspect which improves the functionality or competitiveness of
the brain or the researcher.   
Models and algorithms for learning in these contexts 
have been studied, in great detail, in neuroscience, image processing,
and many other fields, for quite some time; consider e.g. 
the introduction of Boltzmann machines more than a quarter of a century ago \cite{Hinton83}, 
and the now more than ten-year old monograph on learning in graphical models \cite{Jordan98}.  

As one could have expected, there is a trade-off between the complexity of the model to
be learned and the efficiency of the learning algorithms. Boltzmann machines are,
in principle, able to learn very complex models, and provably so, but convergence
is then quite slow; modern applications of those methods
center on various restricted Boltzmann machine models, which can be learned faster \cite{Hinton07}.

When we consider very large systems and/or problems where convergence time of the learning
is a serious concern, then we are primarily interested in those 
very fast processes of learning which could be nick-named
``immediate understanding'', or ``iterative understanding''. By this we mean that
the outcome of learning should be read out directly by one or a series of mathematical transformations
of the data. The central issue is then obviously the accuracy of the learning outcomes. 
In this paper we have revisited some classical and some more recent
algorithms of this kind that take their inspiration from statistical physics. 
The two simplest algorithms considered are {\it naive mean-field} and
{\it independent pair approximation}. As we have shown for neural data,
both of these perform poorly except for small systems (small $N$), of for systems which
have little variability (small value of $N\delta$), where $\delta$ here can be thought of
as a proxy for the deviation from a uniform state). On the other hand, one
generalization of each method, the TAP approximation
for the naive mean-field and the recent
Sessak-Monasson approximation for the independent pair approximation,
perform much better. We believe that there is scope for further ideas, and note particularly the
recent reported generalization of the TAP equations to learning within
a Bethe-lattice approximation; this approach, so far only carried out for pairwise binary models,
could be a promising avenue for learning more generally in large systems where 
an underlying connectivity is locally tree-like \cite{MoraPhD07}.
We also studied the quality of pairwise models using a variety of methods.
We derived mathematical expressions relating the quality of pairwise models 
to the size of the population and the lower order statistics of the spike trains employing 
a number of approximation schemes. 

To conclude, we have here framed the presentation in terms of inferring representations
of neural data and assessing the goodness of the model. Although our focus was
 on neural data analysis, it is important to note that similar problems appear in many other fields of modern biology,
for instance, e.g. in network reconstruction of gene regulatory networks,
in genomic assembly in metagenomics projects, and in many other problems.
Given the present explosion in sequencing technologies it is conceivable 
that the more novel applications will soon appear outside neuroscience.

\bigskip

\bigskip

\noindent {\LARGE \bf Appendices}

\bigskip

\appendix

\renewcommand{\theequation}{A-\arabic{equation}}
\setcounter{equation}{0}
\section{The partition function, entropy and moments of a Gibbs distribution in the limit $N\delta \rightarrow 0$}
 \label{Ising-Low}
 
Suppose we have a true distribution of the following form
\begin{equation}
p(\bs)=\frac{1}{Z}\exp\left[\sum_i h_i s_i + \sum_{i<j} J_{ij} s_i s_j + \sum_{i<j<k}K_{ijk}  s_i s_j s_k +\dots \right].\label{def-third}
\end{equation}
In this Appendix, we first find the relation between the the external fields and pairwise and third order couplings. As we 
show below, this allows us to compute the probability of synchronous spikes. We also compute the entropy 
of this distribution, in the small spike probability regime to derive the expression in Eqs.~\eqref{ents-pert}. This task can 
be accomplished much more easily if we rewrite the distribution of Eq.~ \eqref{def-third} in terms of zero-one variables $r_i$, instead 
of the the spin variables $s_i$, i.e.
\begin{subequations}
\begin{align}
&p(\br)=\frac{1}{Z_p}\exp\left[\beta \Big( \sum_i \calH_i r_i + \sum_{i<j} \calJ_{ij} r_i r_j + \sum_{i<j<k}\calK_{ijk} r_i r_j r_k +\dots \Big) \right]\label{def-third-r}\\
&r_i=(1+s_i)/2,
\end{align}
\label{p-third}
\end{subequations}
\noindent where the auxiliary inverse temperature $\beta$ is introduced because it will allow us to 
compute the entropy, as will become clear later. 

The crucial step in computing the relation between the moments, parameters and entropy
of a distribution is computing its partition function. To compute the partition function of Eq.~\eqref{def-third-r}, $Z$, we 
note that
\begin{eqnarray}
\Gamma_p\equiv \frac{Z}{Z_0}-1&=& \left \langle \exp\left[\sum_{i<j} \beta \calJ_{ij} r_i r_j + \sum_{i<j<k} \beta \calK_{ijk}  r_i r_j r_k +\dots \right]\right \rangle_0-1\nonumber\\
&=&\sum^{\infty}_{n=1}\frac{1}{n!}\left \langle \left[\sum_{i<j} \beta \calJ_{ij} r_i r_j + \sum_{i<j<k} \beta \calK_{ijk}  r_i r_j r_k +\dots \right]^n\right \rangle_0,
\label{Gamma-def}
\end{eqnarray}
where $\langle \rangle_0$ indicates averaging with respect to the distribution 
\begin{equation}
p_0(r)=\frac{1}{Z_0}\exp\left[\sum_i \calH_i r_i \right].
\end{equation}
Note that $p_0$ is {\em not} the independent model for $p$ in Eq.~\eqref{p-third} but only the part of this distribution that includes the fields. In fact, 
as we show in the following (Eq.~\eqref{MH}) the fields of the independent model to $p$ only match $\calH_i$ to $\order(N\delta)$ corrections,
i.e. $\calH_i=\calH^{\rm ind}_i+\order(N\delta)$. When dealing with corrections to the fields and couplings, this note will be important.

Since $\langle r^n_i\rangle_0=\langle r_i \rangle_0\equiv \delta_i $, a term with $l$ distinct indices 
in the expansion of the term inside the average in Eq.~\eqref{Gamma-def} 
is of the order of  $\delta^l$. Therefore, to $\order((N\delta)^3)$, we have
\begin{eqnarray}
\Gamma&=&\sum_{i<j}\sum_{n}\frac{\beta^n}{n!}  J^n_{ij} \delta_i \delta_j \nonumber\\
&+&\sum_{i<j<k}\sum_{n_1,n_2}\frac{\beta^{n_1+n_2}}{n_1! n_2!} [J_{ij}^{n_1} J_{ik}^{n_2}+J_{ij}^{n_1}  J_{jk}^{n_2}+J_{ik}^{n_1} J_{jk}^{n_2}]\delta_i \delta_j \delta_k + \sum_{i<j<k}\sum_{n_1,n_2,n_3}\frac{\beta^{n_1+n_2+n_3}}{n_1! n_2!n_3!} J_{ij}^{n_1} J_{ik}^{n_2} J_{jk}^{n_3} \delta_i \delta_j \delta_k \nonumber\\
&+&\sum_{i<j<k} \sum_{n=1}\frac{\beta^n}{n!} K^{n}_{ijk}  \delta_i \delta_j \delta_k  + \sum_{i<j<k}\sum_{n_1,n_2}\frac{\beta^{n_1+n_2}}{n_1! n_2!}K^{n_1}_{ijk}[J^{n_2}_{ij}+J^{n_2}_{ik}+J^{n_2}_{jk}]  \delta_i \delta_j \delta_k \nonumber\\
&+&\sum_{i<j<k}\sum_{n_1,n_2,n_3}\frac{\beta^{n_1+n_2+n_3}}{n_1! n_2!n_3!}K^{n_1}_{ijk}[J^{n_2}_{ij}J^{n_3}_{ik}+J^{n_2}_{ij}J^{n_3}_{jk}+J^{n_2}_{ik}J^{n_3}_{jk}]  \delta_i \delta_j \delta_k\nonumber\\
&+&\sum_{i<j<k}\sum_{n_1,n_2,n_3,n_4}\frac{\beta^{n_1+n_2+n_3+n_4}}{n_1! n_2!n_3!n_4!}K^{n_1}_{ijk}J^{n_2}_{ij}J^{n_3}_{ik} J^{n_3}_{jk}  \delta_i \delta_j \delta_k + \order((N\delta)^4),
\end{eqnarray}
where the sums over $n,n_1,n_2,n_3$ and $n_4$ run from $1$ to infinity. Performing these sums yields
\begin{eqnarray}
\Gamma&=&\sum_{i<j}\phi_{ij}\delta_i \delta_j +\sum_{i<j<k} [\phi_{ij}\phi_{ik}+\phi_{ij}\phi_{jk}+\phi_{ik}\phi_{jk}+\phi_{ij}\phi_{ik}\phi_{jk}] \delta_i \delta_j\delta_k\nonumber\\
&+&\sum_{i<j<k} \phi_{ijk}(1+\phi_{ij})(1+\phi_{ik})(1+\phi_{jk}) \delta_i \delta_j\delta_k +\order((N\delta)^4)
\label{Gamma-Ndelta4}
\end{eqnarray}
where $\phi_{ij}=\exp(\beta \calJ_{ij})-1$ and $\phi_{ijk}=\exp(\beta \calK_{ijk})-1$.
From Eq.~\eqref{Gamma-Ndelta4}, we can immediately compute the relation between the means, pairwise and three-point correlation functions and the parameters of
the distribution. For $\beta=1$, we have 
\begin{subequations}
\begin{align}
&\tim_i\equiv\langle r_i \rangle=\frac{\partial \log Z}{\partial \calH_i}=\delta_i+\frac{\partial \log(1+\Gamma)}{\partial \calH_i}=\delta_i \left[1+\sum_{j\neq i} \phi_{ij} \delta_j +\order((N \delta)^2)\right]\label{MH}\\
&\tiC_{ij}\equiv\langle r_i r_j \rangle=\frac{\partial \log Z}{\partial \calJ_{ij}}=\frac{\partial \log (1+\Gamma)}{\partial J_{ij}}=\exp(\calJ_{ij})\delta_i \delta_j\left[1+\sum_{k\neq i, j} [\phi_{ik}+\phi_{jk}+\phi_{ik}\phi_{jk}]\delta_k+\order((N\delta)^2)\right]\label{CJ}\\
&\tiC_{ijk}\equiv\langle r_i r_j r_k  \rangle=\frac{\partial \log Z}{\partial \calK_{ijk}}=\frac{\partial \log (1+\Gamma)}{\partial \calK_{ijk}}=\exp(\calK_{ijk})\exp(\calJ_{ij}+\calJ_{ik}+\calJ_{jk})\delta_i \delta_j \delta_k+\order(N\delta^4).\label{C3H}
\end{align}
\label{moments-couplings}
\end{subequations}
\noindent The relations between means and pairwise correlations and the external fields and 
pairwise couplings in Eq.~\eqref{moments-couplings} a and b to their leading orders were reported previously 
in \cite{Roudi09}, using a slightly different approach. However, the corrections and three-neuron 
correlations were not computed there.

An interesting result of this calculation is  a relation between the three-neuron correlations for the 
pairwise distribution, i.e. when $\calK_{ijk}=0$,  and the lower moments
\begin{equation}
\tiC^{\rm pair}_{ijk}\equiv\langle r_i r_j r_k  \rangle_{\rm pair}=\frac{\tiC_{ij}\tiC_{ik}\tiC_{jk}}{\tim_i\tim_j\tim_k}+\order(N\delta^4).
\label{third-pair}
\end{equation}
The fact that, to leading order in $N\delta$, the external fields and couplings are determined by means 
and pairwise correlations allows us to compute the leading-order probabilities of synchronous 
spikes reported as we did in sec. \ref{sec:synch-spikes}.

We can now use Eq.~\eqref{Gamma-Ndelta4} to find the entropies of the distribution in Eq.~\eqref{def-third-r}
\begin{equation}
S=\log(Z)-\frac{\partial \log Z}{\partial \beta}=\log(Z_0)+\log(1+\Gamma)-\sum_{i}\calH_i \langle r_i \rangle -\sum_{i<j} \calJ_{ij}  \langle r_i  r_j \rangle -  \sum_{i<j<k} \calK_{ijk} \langle r_i  r_j r_k \rangle.
\end{equation}
From Eqs.\eqref{moments-couplings}, for an independent fit to the distribution Eq.~\eqref{def-third-r}, we have $\calH^{\rm ind}_i=\calH_i$. Consequently,
\begin{eqnarray}
S_{\rm ind}-S&=&-\Gamma+\sum_{i<j} \calJ_{ij}  \langle r_i r_j \rangle+\sum_{i<j<k} \calK_{ijk} \langle  r_i  r_j r_k \rangle +\order((N\delta)^4)\nonumber\\
&=&-\sum_{i<j} \phi_{ij} \tim_i \tim_j+\sum_{i<j} \calJ_{ij}  \langle r_i r_j \rangle +\order((N\delta))^3.
\end{eqnarray}
Using the fact that $\tim_i = \delta_i+ \order(N\delta^2)$ from Eq.~\eqref{MH}, and $\phi_{ij}=\exp(\calJ_{ij})-1=\tiC_{ij}/(\tim_i\tim_{j})-1+\order(N\delta)$ from
Eq.~\eqref{CJ}, we get Eq.~\eqref{ents-perta} with $\gind$ defined
as
\begin{subequations}
\begin{align}
&\gind\equiv \frac{1}{N(N-1)}\sum_{i<j} [(1+\rho_{ij}) \log(1+\rho_{ij})-\rho_{ij}] \frac{\tim_i}{\delta}  \frac{\tim_j}{\delta} \label{gind} \\
&\rho_{ij}=\frac{\tiC_{ij}}{\tim_i \tim_j}-1.
\end{align}
\end{subequations}

For a pairwise model to the distribution in Eq.~\eqref{def-third-r}, we have $\calH^{\rm pair}_i=\calH_i+\order((N\delta)^2)$ and $\calJ^{\rm pair}_{ij}=\calJ_{ij}+\order((N\delta)^2)$ and thus
\begin{equation}
S_{\rm pair}-S=-\sum_{i<j<k} \phi_{ijk}(1+\phi_{ij})(1+\phi_{ij})(1+\phi_{ij})\delta_i \delta_j\delta_j+\sum_{i<j<k} \calK_{ijk} \langle r_i r_j r_k \rangle.
\end{equation}
Using Eq.~\eqref{C3H} and Eq.~\eqref{third-pair}, this will lead to Eq.~\eqref{ents-perta} with $\gpair$ defined as

\begin{subequations}
\begin{align}
&\gpair \equiv \frac{1}{N(N-1)(N-2)}\sum_{i<j} \Big( (1+\rho_{ijk}) [\log(1+\rho_{ij})-\log(1+\rho^{\rm pair}_{ijk})]-[\rho_{ijk}-\rho^{\rm pair}_{ijk}] \Big) \frac{\tim_i}{\delta}  \frac{\tim_j}{\delta}  \frac{\tim_j}{\delta} \label{gpair} \\
&\rho_{ijk}=\frac{\tiC_{ijk}}{\tim_i \tim_j\tim_k}-1, \  \  \  \ \rho^{\rm pair}_{ijk}=\frac{\tiC^{\rm pair}_{ijk}}{\tim_i \tim_j\tim_k}-1.
\end{align}
\end{subequations}

\renewcommand{\theequation}{B-\arabic{equation}}
\setcounter{equation}{0}
\section{The independent-pair approximation for the external fields in the limit $\delta \to 0$}
\label{lowrate-pairs}

Replacing $\delta_i=\exp(h_i)/(1+\exp(h_i))$ in Eq.~\eqref{MH} and solving to find $h_i$ we get
\begin{eqnarray}
\calH_i&=&-\log\left[\frac{1-\langle r_i \rangle}{\langle r_i \rangle}\right]-\sum_{j\neq i}\phi_{ij} \delta_j+\order((N\delta)^2)\nonumber\\
&=&-\log\left[\frac{1-\langle r_i \rangle}{\langle r_i \rangle}\right]-\sum_{j\neq i}\frac{\langle r_i r_j \rangle-\langle r_i\rangle \langle r_j \rangle}{\langle r_i\rangle}+\order((N\delta)^2),
\end{eqnarray}
where in the second line we have used the fact that $\phi_{ij}=\frac{\langle r_i r_j \rangle}{\langle r_i\rangle \langle r_j\rangle}-1+\order(N\delta)$ from 
Eq.~\eqref{CJ} and that $\langle r_i \rangle=\delta_i+\order(N\delta)$. Changing the variables from $r_i=0,1$ to the original spin variables $s_i=\pm 1$,
we have
\begin{equation}
h_i=\frac{\calH_i}{2}+\sum_{j\neq i} \frac{\calJ_{ij}}{4}=\frac{1}{2}\log\left[\frac{1+m_i}{1-m_i}\right]-\sum_{j\neq i} \frac{C_{ij}}{4(1+m_i)}+\sum_{j\neq i} J_{ij}.
\label{corrections-h}
\end{equation}
For the IP approximation, on the other hand, we have
\begin{equation}
h^j_i=J_{ij}+\frac{1}{2}\log\left[\frac{1+m_i}{1-m_i}\right]-\frac{C_{ij}}{4(1+m_i)}.
\label{corrctions-hji}
\end{equation}
Summing $h^j_i$ over $j$ and subtracting the over counted terms from single spin contributions 
gives the same expression as Eq.~\eqref{corrections-h}. Therefore, the IP approximation to the external fields 
get the leading term and the first order corrections of the low-rate approximation correctly.

\renewcommand{\theequation}{C-\arabic{equation}}
\setcounter{equation}{0}
\section{Independent-triplet (IT) approximation}
\label{triplet-expansion}

Considering the following Gibbs distribution over three boolean variables $i$, $j$, and $k$,
\begin{equation}
p_{ijk}(r_i,r_j,r_k)=\frac{1}{Z_{ijk}} \exp\Big[\calH_i r_i+\calH_j r_j +\calH_k r_k+\calJ_{ij} r_i r_j+\calJ_{ik} r_i r_k + \calJ_{jk} r_j r_k\Big],
\end{equation}
we can use the definition of the means and correlations and write
\begin{equation}
\left( \begin{array}{c}
\tim_i\\
\tim_j\\
\tim_k\\
\tiC_{ij}\\
\tiC_{ik}\\
\tiC_{jk}\\
\tiC_{ijk}\\
1
\end{array} \right)
=
\left( \begin{array}{cccccccc}
0&0&0&0&1&1&1&1\\
0&0&1&1&0&0&1&1\\
0&1&0&1&0&1&0&1\\
0&0&0&0&0&0&1&1\\
0&0&0&0&0&1&0&1\\
0&0&0&1&0&0&0&1\\
0&0&0&0&0&0&0&1\\
1&1&1&1&1&1&1&1
\end{array} \right)
\left( \begin{array}{c}
p_{ijk}(0,0,0)\\
p_{ijk}(0,0,1)\\
p_{ijk}(0,1,0)\\
p_{ijk}(0,1,1)\\
p_{ijk}(1,0,0)\\
p_{ijk}(1,0,1)\\
p_{ijk}(1,1,0)\\
p_{ijk}(1,1,1)
\end{array} \right).
\end{equation}
Inverting the matrix of coefficients we can express the probabilities in terms of the means, pairwise and third order correlations. Using the result in
\begin{subequations}
\begin{align}
&\calH^{jk}_i=\log\left[\frac{p_{ijk}(1,0,0)}{p_{ijk}(0,0,0)}\right]\\
&\calJ^{k}_{ij}=\log\left[\frac{p_{ijk}(0,0,0)p_{ijk}(1,1,0)}{p_{ijk}(1,0,0)p_{ijk}(0,1,0)}\right]
\end{align}
\label{triplets-params}
\end{subequations}
and similar equations for the other fields and couplings, we can also express these parameters in terms of means, pairwise and third order correlations.
Before using Eqs.~\eqref{triplets-params} as approximations for the parameters of a pairwise model, we need to perform two more steps. 

The first step is a familiar one that we noted when dealing with the external fields in the independent-pair approximation, namely the
fact that $\calH^{jk}_i$ depends on $j$ and $k$ in addition to $i$ and that $\calK^{k}_{ij}$ depends on $k$. We can use the same logic that we used 
in building an independent-pair approximation to the external fields, and build  an approximation to the external fields 
that does not depended on $j$ and $k$, and an approximation to the couplings that does not depend on $k$. For example, for the case 
of the pairwise couplings we get
\begin{equation}
\calJ^{\rm IT}_{ij}=\calJ^{\rm IP}_{ij}+\sum_{k\neq i,j} \log\left[\frac{(1-\tiC_{ijk}/\tiC_{ij}) (1- (\tim_k-\tiC_{ik}-\tiC_{jk}+\tiC_{ijk}) / (1-\tim_i-\tim_j-\tiC_{ij}) )  }   { (1-(\tiC_{ik}-\tiC_{ijk}) / (\tim_i-\tiC_{ij}) ) ( 1-(\tiC_{jk}-\tiC_{ijk}) /(\tim_j-\tiC_{ij}) ) }\right]
\label{triplets-J}
\end{equation}

The second step has to do with the fact that Eqs.~\eqref{triplets-params} (as well as their transformed version after performing the first step, e.g. Eq.~\ref{triplets-J}) depend on the
third order correlations in addition to the pairwise correlations and the means. Hence to derive an expression that 
relates model parameters to pairwise correlations and means we should first find the third order correlation in terms of them.
Note that this step is not present in the independent-pair approximation. To express the third order correlations in terms of
the lower order statistics we take advantage of the following equation
\begin{equation}
p_{ijk}(0,0,1)p_{ijk}(0,1,0)p_{ijk}(1,0,0)p_{ijk}(1,1,1)=p_{ijk}(0,0,0)p_{ijk}(0,1,1)p_{ijk}(1,0,1)p_{ijk}(1,1,0).
\label{triplets-lower}
\end{equation}
Writing the probabilities in terms of the moments, this equation can be solved to find the third order correlations in terms of the means and pairwise correlations.
The equations have two imaginary solutions for $\tiC_{ijk}$, which are unphysical, and one real solution, which is
the correct solution to be considered. The resulting expression for $\tiC_{ijk}$ in terms of the means and pairwise correlations
is complicated. However, in the limit of $\tim_i,\tim_j,\tim_k\rightarrow 0$, it can be shown to have the same form as the one reported
in Eq.~\eqref{third-pair}. We noted in the text that in this limit, the IP approximations to the couplings will give the same result
as the leading order term of Eq.~\eqref{CJ} for the couplings. With the independent-triplet approximation, we can go
one step further, and as can be shown by doing a small amount of algebra, we can recover $\order(N\delta)$  corrections
to the couplings in that we calculated in Eq.~\eqref{CJ}.

As mentioned in sec. \ref{IPA-EX}, one can continue the above process to build approximations based on quadruples of spins and so on. 
However, this soon becomes difficult in practice for the reason that solving equations of the type Eq.~\eqref{triplets-lower} to find 
the higher moments in terms of the means and pairwise correlations will be as difficult as the original problem of finding the 
external fields and couplings of the original $N$ body problem in terms of the means and correlations. Nevertheless,
this simple triplet expansion offer an alternative derivation of Eqs.~\eqref{ents-pert}, simpler than the derivations in \cite{Roudi09} and 
Appendix \ref{Ising-Low}. Here we show this for Eq.~\eqref{ents-pertb}, as deriving
Eq.~\eqref{ents-perta} will be similar but less invoved.To derive Eq.~\eqref{ents-pertb} using the triplet expansion, we
 approximate the entropy of the whole system of $N$ neurons
as a sum of the entropies of all triplets denoted by $S_{ijk}$. We then expand the resulting expression 
keeping terms of up to $\order(\delta^3)$ noting that $\tiC_{ijk}\sim \order(\delta^3)$, and $\tiC_{ij}\sim \order(\delta^2)$. The result takes the form of
\begin{eqnarray}
S&=&\sum_{i<j<k} S_{ijk}=-\sum_{i<j<k}\sum_{s_i,s_j,s_k} p_{ijk}(s_i,s_j,s_k) \log(p_{ijk}(s_i,s_j,s_k))\nonumber\\
&=&-\sum_{i<j<k}Q(\tiC_{ijk})-\tiC_{ijk}\nonumber\\
&-&\sum_{i<j<k} \tiC_{ijk} \left[-\log(\tiC_{ij})-\log(\tiC_{jk})-\log(\tiC_{ik})+\log(\tim_i)+\log(\tim_j)+\log(\tim_k)\right] \nonumber\\
&-&\sum_{i<j<k} Q(\tim_i-\tiC_{ij}-\tiC_{jk}) +Q(\tim_i-\tiC_{ij}-\tiC_{jk})+ Q(\tim_i-\tiC_{ij}-\tiC_{jk})\nonumber\\
&-&\sum_{i<j<k}Q(\tiC_{ij})+Q(\tiC_{ik})+Q(\tiC_{jk})\nonumber\\
&-&\sum_{i<j<k} Q(1-\tim_i-\tim_j-\tim_k+\tiC_{ij}+\tiC_{ik}+\tiC_{jk})
\label{tiplet-ent-third}
\end{eqnarray}
where $Q(x)=x\log(x)$. For a pairwise model, the independent-triplet approximation to the entropy in the limit $\tim\to 0$ has the same form, except that $\tiC_{ijk}$ of 
the true model should be replaced by $\tiC^{\rm pair}_{ijk}=(\tiC_{ij}\tiC_{ik})\tiC_{jk})(\tim_i\tim_j\tim_k)^{-1}$ (see Eq.~\eqref{third-pair}), i. e.
\begin{eqnarray}
S_{\rm pair}=&-&\tiC^{\rm pair}_{ijk}\nonumber \\
&-&\sum_{i<j<k} Q(\tim_i-\tiC_{ij}-\tiC_{jk}) +Q(\tim_i-\tiC_{ij}-\tiC_{jk})+ Q(\tim_i-\tiC_{ij}-\tiC_{jk})\nonumber\\
&-&\sum_{i<j<k}Q(\tiC_{ij})+Q(\tiC_{ik})+Q(\tiC_{jk})\nonumber\\
&-&\sum_{i<j<k} Q(1-\tim_i-\tim_j-\tim_k+\tiC_{ij}+\tiC_{ik}+\tiC_{jk})
\label{tiplet-ent-pair}
\end{eqnarray}
Using Eqs.~\eqref{tiplet-ent-third}, \eqref{tiplet-ent-pair} and \eqref{C3H} yields Eq.~\eqref{ents-pertb}.

\renewcommand{\theequation}{D-\arabic{equation}}
\setcounter{equation}{0}
\section{Derivation of TAP equations from Belief Propagation}
\label{TAP-BP}
In this appendix we derive the TAP equations (Eq.~\eqref{TAP}) starting
from the Belief Propagation update rules.  Let us begin with
the result to be established. The TAP equations are
a set of nonlinear equations for the magnetizations $m_i$, which
we will write:
\begin{equation}
\tanh^{-1} m_i = h_i + \sum_{j\in \partial_i} 
\left( \epsilon J_{ij}m_j - \epsilon^2 J_{ij}^2 m_i(1-m_j^2)\right) + 
 {\cal O}(\epsilon^3).
\label{eq:TAP-with-epsilon}
\end{equation}
Here $h_i$ is the external field 
acting on spin $i$, $\epsilon J_{ij}$ are the pairwise couplings, and the notation$j\in \partial_i$ means that 
the sum is over neurons $j$ connected with neuron $i$.
As we also did in the text in Eq.~\eqref{TAP}, the above equation is generally quoted with $\epsilon$
set to one and without the error term of cubic order in $\epsilon$.

Starting from the pairwise distribution Eq.~\eqref{pairwise_dist}, we define the following distribution, with the auxiliary variable
$\epsilon$ that we set to $1$ in the end of our calculation

\begin{equation}
p^{\epsilon}(\bs) =  \frac{1}{Z}   \exp\left[\sum_i h_i s_i + \epsilon \sum_{i<j} J_{ij} s_i s_j\right]
\label{eq:statmodel-with-epsilon}
\end{equation}
and the exact marginal distribution over spin $s_i$ is defined by
\begin{equation}
p^ {\epsilon}_i(s_i) =  \sum_{{\bf s}\setminus s_i}p^{\epsilon}(\bs)
\label{eq:marginal-with-espilon}
\end{equation}
where the sum goes over all spins except $s_i$. The exact magnetization of spin $s_i$ is then
\begin{equation}
m_i = 
  \frac{\sum_{\bf s}s_i\exp\left(\sum_i h_i s_i + \epsilon \sum_{i<j} J_{ij} s_i s_j\right)}
       {\sum_{\bf s}   \exp\left(\sum_i h_i s_i + \epsilon \sum_{i<j} J_{ij} s_i s_j\right)}.
\label{eq:m_i-with-epsilon}
\end{equation}

Belief Propagation is a family of methods for approximately computing the marginal distributions
of probability distributions \cite{Kschischang2001,YedFreWei2003,MezardMontanari2009}. Since the 
model defined by Eq.~\eqref{eq:statmodel-with-epsilon} contains only pairwise interactions, it is 
convenient to adopt the pairwise Markovian Random Field formalism of Yedidia, 
Freeman and Weiss \cite{YedFreWei2003}.  Note that the important recent contribution by M\'ezard and Mora uses
a more general formalism, which may prove to be more convenient in the perspective of extending an "inverse BP" learning algorithm
beyond pairwise models \cite{MoraPhD07}.

Belief Propagation applied to the model in Eq.~\eqref{eq:statmodel-with-epsilon} in the Yedidia-Weiss-Freeman 
formalism is built on probability distributions, called BP messages, associated with every 
directed link in the graph. If $i\to j$ is such a link, starting at $i$ and ending at $j$, then the BP 
message $\eta_{i\to j}(s_j)$ is a probability distribution on the variable $s_j$ associated to node 
where the link ends. For Ising spins one can use the parametrization
\begin{equation}
\eta_{i\to j}(s_j) = \frac{1+s_jm_{i\to j}}{2},
\label{eq:m-definition}
\end{equation}
where $m_{i\to j}$ is a real number called the cavity magnetization. BP is characterized
by two equations, the {\it Belief Propagation update equations}, and the {\it Belief Propagation output equations}.  The BP update 
equations are used iteratively to find a fixed point, which is an extremum of the Bethe free energy.  At the
fixed point, the BP update equations form a (large) set of compatibility conditions for the $\eta$'s which, for the model Eq.~\eqref{eq:statmodel-with-epsilon}, read
\begin{equation}
\eta_{j\to i}(s_i) = \frac{1}{\Omega_{j\to i}}
                     \sum_{s_j} e^{h_j s_j + \epsilon J_{ij} s_i s_j}
                     \prod_{k\in \partial_j\setminus i} \eta_{k\to j}(s_j).
\label{eq:BP-update}
\end{equation}
The BP output equations determine the marginal probability distributions from the $\eta$'s and read
\begin{equation}
p^{\epsilon}_{i}(s_i) = \frac{1}{\Omega_{i}}
                     e^{h_i s_i} \prod_{j\in \partial_i} \eta_{j\to i}(s_i),
\label{eq:BP-output}
\end{equation}
where $\Omega_{j\to i}$ and $\Omega_{i}$ in Eqs.~\eqref{eq:BP-update} and \eqref{eq:BP-output} are normalizations. To lighten the notation, we will
not distinguish between the exact marginals, as in Eq.~\eqref{eq:marginal-with-espilon}, and the 
approximate marginals from BP, as in Eq.~\eqref{eq:BP-output}. 

We now write the BP update and BP output equations using the cavity 
magnetizations, $m_{i\to j}$, from Eq.~\eqref{eq:m-definition}. The notation simplifies if one define an ancillary quantity
\begin{equation}
q_{i\to j} = \frac{1}{2}\log\frac{1+m_{i\to j}}{1-m_{i\to j}} = \tanh^{-1}m_{i \to j}
\label{eq:q-definition}
\end{equation}
in terms of which the BP update equation can be written as
\begin{equation}
m_{j\to i} = \tanh(\epsilon J_{ij})
\tanh(h_j + \sum_{k\in \partial_j\i} q_{k\to j})
\label{eq:BP-update-with-m-and-q}
\end{equation}
and the BP output equation as
\begin{equation}
m_{i} = \tanh(h_i + \sum_{j\in i} q_{j\to i}).
\label{eq:BP-output-with-m-and-q}
\end{equation}
The task is now to expand the right hand side of Eq.~\eqref{eq:BP-output-with-m-and-q} in $\epsilon$
and compare with the TAP equations, Eqs.~\eqref{eq:TAP-with-epsilon}. To do this, we first note that
\begin{equation}
q_{j\to i} = \epsilon J_{ij} 
\tanh(h_j + \sum_{k\in \partial_j\i} q_{k\to j}) + 
{\cal O}(\epsilon^3)= \epsilon J_{ij} m_{j}+\order(\epsilon^3).
\label{eq:epsilon-expansion-1}
\end{equation}
This follows from expanding Eq.~\eqref{eq:BP-update-with-m-and-q} in $\epsilon$, using the result
in Eq.~\eqref{eq:q-definition}, and finally, expanding the logarithm. We then rewrite the BP output equation
\eqref{eq:BP-output-with-m-and-q} as 
\begin{equation}
\tanh^{-1}m_{i} = h_i + \epsilon\sum_{j\in i} J_{ij}
\tanh(h_j - q_{i\to j}+ \sum_{k\in \partial_j} q_{k\to j})
          + \order(\epsilon^3).
\label{eq:epsilon-expansion-3}
\end{equation}
Since $m_j = \tanh(h_j + \sum_{k\in \partial_j} q_{k\to j})$ (no expansion in $\epsilon$) we want to
separate $q_{i\to j}$ and $h_j +\sum_{k\in \partial_j} q_{k\to j}$ in the arguments of 
the $\tanh$'es in Eq.~\eqref{eq:epsilon-expansion-3}, and according to Eq.~\eqref{eq:epsilon-expansion-1}, $q_{i\to j}$ is of
order $\epsilon$. This means that we can write, to order $\epsilon$,
\begin{equation}
\tanh(h_j - q_{i\to j}+ \sum_{k\in \partial_j} q_{k\to j})= m_j -\epsilon J_{ij}m_i (1-m_j^{2}).
\end{equation}
Introducing this into Eq.~\eqref{eq:epsilon-expansion-3}, we finally have 
\begin{equation}
\tanh^{-1}m_{i} = h_i + \sum_{j\in i} 
(\epsilon J_{ij}m_j -\epsilon^2 J_{ij}^2m_i (1-m_j^{2}))
          + \order(\epsilon^3)
\label{eq:epsilon-expansion-4}
\end{equation}
which was to be proved.
\bibliography{mybibliography}
\end{document}